\documentstyle[12pt]{article}
\begin{document}
\title{The BFT Method With Chain Structure}
\author{A. Shirzad$^{a,b}$\thanks{shirzad@ipm.ir} \\ M. Monemzadeh$^{a}$\thanks{e-mail: monemzadeh@sepahan.iut.ac.ir} \\  \\
  $^a$~{\it Department of  Physics, Isfahan University of Technology (IUT)}\\
{\it Isfahan,  Iran,} \\
 $^b$~{\it Institute for Studies in Theoretical Physics and Mathematics (IPM)}\\
{\it P. O. Box: 19395-5531, Tehran, Iran.}}
\date{}
\maketitle

\begin{abstract}
We have constructed a modified BFT method that preserves the chain
structure of constraints. This method has two advantages: first,
it leads to less number of primary constraints such that the
remaining constraints emerge automatically; and second, it gives
less number of independent gauge parameters. We have applied the
method for bosonized chiral Schwiger model. We have constructed a
gauge invariant embedded Lagrangian for this model.
\end{abstract}
Dirac as a pioneer, quantized classical gauge theories by
converting Poisson brackets to quantum commutators \cite{Dirac}.
However, for second class constraint systems it is necessary to
replace Poisson brackets by Dirac brackets and then convert them
to quantum commutators. Sometimes this process implies problems
such as factor ordering which makes this approach improper. The
BFT method, however, solves this ambiguity by embedding the phase
space in a larger space including some auxiliary fields
\cite{BatFrad,BatTyu}. In this way one can convert second class
constraints to first class ones and then apply the well-known
quantization method of gauge theories \cite{BanBanGh1,BanBanGh2}.
In our previous paper \cite{MonemShir} we showed that if one
chooses arbitrary parameters of the BFT method suitably then the
power series of auxiliary fields for the embedded constraints and
Hamiltonian could be truncated in some cases.

In this paper we want to preserve the {\it chain structure} of a
second class system (except for the last element of the chain)
during the BFT embeding.
 The main idea of the chain structure, as fully discussed
in \cite{LoranShir}, is that it is possible to derive the
constraints as commuting distinct chains such that within each
chain the following iterative relation holds
 \begin{equation}
\Phi_\alpha^a=\{\Phi_{\alpha-1}^a , H_c\},
 \label{b1}
 \end{equation}
where $\Phi_0^a$ stand for primary constraints. The advantages of
this method will be discussed afterward.

Consider a second class constraint system described by the
Hamiltonian $H_0$ and a set of second class constraints
$\Phi_\alpha ;\hspace {3mm}\alpha=1,...N$ satisfying the algebra
\begin{equation}
 \Delta_{\alpha\beta}=\{\Phi_\alpha , \Phi_\beta\}
 \label{a1}
 \end{equation}
where $\Delta$ is an antisymmetric and invertible matrix. For
simplicity and without loss of generality we suppose that the
second class constraints $\Phi_\alpha$ are elements of one chain.
The results can be extended to multi-chain system by adding the
chain index (like the superscript $a$ in Eq. (\ref{b1})) to the
constraints $\Phi_\alpha$. For converting this second class system
into a gauge system, one can enlarge the phase space by
introducing auxiliary variables $\eta$ where we {\it assume} their
algebra to be
 \begin{equation}
 \omega^{\alpha\beta}=\{\eta^\alpha , \eta^\beta\}.
 \label{a2}
 \end{equation}
We demand that the embedded constraints $\tau_\alpha(q,p,\eta)$
and Hamiltonian $\tilde{H}(q,p,\eta)$ in the extended phase space
satisfy the following algebra

 \begin{eqnarray}
 \{\tau_\alpha , \tau_\beta\}&=&0
 \label{a4}
 \\ \{\tau_\alpha , \tilde{H}\}&=&\tau_{\alpha+1} \hspace{1.5cm}
 \alpha=1,...N-1
 \label{a14}
 \\
  \{\tau_N , \tilde{H}\}&=&0.
 \label{h1}
   \end{eqnarray}
This gives an abelian first class chain such that its terminating
element commute with the Hamiltonian. We call this system a {\it
semi-strongly involutive system}; compared with strongly
involutive one in which the constraints commute with the
Hamiltonian.

As discussed in \cite{BatFrad, BatTyu, BanBanGh1}, considering the
power series
 \begin{equation}
 \tau_\alpha=\sum_{n=0} ^{\infty}\tau_\alpha^{(n)} \hspace {2cm}
 \tau_\alpha^{(n)}\sim\eta^n
 \label{a5}
 \end{equation}
\begin{equation}
 \tilde{H}=\sum_{n=0}^\infty H^{(n)}
 \hspace{2cm}H^{(n)}\sim\eta^n.
 \label{a15}
 \end{equation}
 in which $\tau_\alpha^{(0)}=\Phi_\alpha$ and $H^{(0)}=H_c(q,p)$,
 one can show that these may be solutions to Eqs. (\ref{a4}) and (\ref{a14})
 if
\pagebreak
\begin{eqnarray}
 \tau_\alpha ^{(1)}&=&\chi_{\alpha\beta}(q,p) \eta^\beta
 \label{a11}
\\
  \tau_\alpha ^{(n+1)}&=&-\frac{1}{n+2} \eta^\beta
 \omega_{\beta\gamma}\chi^{\gamma\delta}B_{\delta\alpha}^{(n)}; \hspace {1cm}n\geq1
 \label{a13}
\\
 H^{(n+1)}&=&-
 \frac{1}{n+1}\eta^\alpha\omega_{\alpha\beta}\chi^{\beta\gamma}\Lambda^{(n)}_\gamma
 \label{a21}
 \end{eqnarray}
in which
\begin{eqnarray}
 B_{\alpha\beta}^{(1)}&=&  \{\tau_{[\alpha}^{(0)} ,
 \tau_{\beta]}^{(1)}\}_{(\eta)}
 \label{a9}
\\
 B_{\alpha\beta}^{(n)}&=&\frac{1}{2}B_{[\alpha\beta]}=\sum_{m=0} ^n \{\tau_\alpha^{(n-m)} ,
 \tau_\beta^{(m)}\}+ \sum_{m=0} ^{n-2}\{\tau_\alpha^{(n-m)} ,
 \tau_\beta^{(m+2)}\}_{(\eta)} \hspace {1cm} n\geq2
 \label{a10}
\\
 \Lambda_\alpha^{(0)}&=&\{\tau_\alpha^{(0)} , H^{(0)}\} -
 \tau_{\alpha+1}^{(0)} \hspace{6.5cm} \alpha<N
 \label{a18}
\\
 \Lambda_\alpha^{(1)}&=&\{\tau_\alpha^{(1)} , H^{(0)}\} + \{\tau_\alpha^{(0)} ,
  H^{(1)}\} + \{\tau_\alpha^{(2)} , H^{(1)}\}_{(\eta)}-
 \tau_{\alpha+1}^{(1)} \hspace{1.2cm} \alpha<N
 \label{a19}
\\
  \Lambda_\alpha^{(n)}&=&\sum_{m=0}^{n}\{\tau_\alpha^{(n-m)} , H^{(m)}\} +
  \sum_{m=0}^{n-2}\{\tau_\alpha^{(n-m)} , H^{(m+2)}\}_{(\eta)} \nonumber \\ & & \hspace{1.5cm}+
 \{\tau_\alpha^{(n+1)} , H^{(1)}\}_{(\eta)}-
 \tau_{\alpha+1}^{(n)} \hspace{1.5cm} n\geq2 \hspace{1.5cm}\alpha<N
 \label{a20}
 \end{eqnarray}
and $\chi_{\alpha\beta}(q,p)$ should satisfy the following
equation
\begin{equation}
 \Delta_{\alpha\beta}+\chi_{\alpha\gamma}\omega^{\gamma\lambda}\chi_{\beta\lambda}=0.
 \label{a12}
 \end{equation}
For $\alpha=N$, due to Eq. (\ref{h1}), the last terms in Eqs.
(\ref{a18}-\ref{a20}) are absent.
 The above results for semi-strongly involutive system is
 different from our previous results for strongly involutive one
 \cite{BatFrad, BatTyu, BanBanGh1} in terms
 $\tau_{\alpha+1}^{(0)}, \tau_{\alpha+1}^{(1)}$ and
 $\tau_{\alpha+1}^{(n+1)}$ in Eqs. (\ref{a18}-\ref{a20}) respectively. We remind that
 \cite{MonemShir}, the aim is to choose a suitable solution for
 $\chi_{\alpha\beta}$ in Eq. (\ref{a12}) such that the series for
 $\tau_\alpha$ and $\tilde{H}$ truncates after a few steps. For
 example when $\Delta$ is the symplectic matrix $J$, the choice $\omega=-J$ and $\chi=J$
 solves(\ref{a12}); similarly when $\Delta$ is a constant
 (antisymmetric) matrix, the choice $\omega=-\Delta$ and $\chi=1$
 is appropriate.

Now let see what is the advantage of the chain structure in our
modified BFT method. We emphasize on two points:

1) Suppose we are given a singular Lagrangian $L$ which leads to
one primary constraint $\Phi_1$. Let the secondary constraints
emerge as a second class chain with elements $\Phi_\alpha;
(\alpha=1,...,N)$ resulting from consistency of the constraints
implicit in the chain relation (\ref{b1}). After embedding one
finds in the traditional BFT method a Hamiltonian $\tilde{H}$
together with $N$ constraints, all in strongly involution. The
constraints can be viewed in this case as $N$ {\it given primary
constraints}. However, preserving the chain structure, one
ultimately obtains an embedded Hamiltonian with just one primary
constraint. The other $N-1$ constraints are then obtained
automatically from the consistency conditions.

It should be added that sometimes it is possible to reconstruct a
singular Lagrangian from a given canonical Hamiltonian and primary
constraints, even though it is not guarantied generally. However,
the less the number of primary constraints, the more the chance to
find the original Lagrangian which gives the desired primary
constraints and Hamiltonian. In the following we will give an
example to show this point. We think that our modified BFT method
improves the chance of finding a corresponding Lagrangian yielding
the embedded primary constraints.

To be more precise, when the Hamiltonian is quadratic and the
 primary constraints are linear with respect to the phase space coordinates,
one can easily reconstruct the corresponding singular Lagrangian.
To do this, one should solve the constraint equations for a number
of momenta; and then insert just linear terms with respect to the
corresponding velocities (with coefficients given by the solutions
of the momenta) in the Lagrangian. The remaining quadratic terms
of the Lagrangian can be found from the corresponding terms of the
Hamiltonian in a regular way.

The important point is that for the cases considered in this paper
($\Delta =J$ or $\Delta=$ constant) the constraints are
necessarily linear before embedding. As explained in more details
in \cite{MonemShir} after embedding, the constraints remain linear
with respect to the coordinates of the extended phase space.
Moreover, if the original Hamiltonian is quadratic, it would
remain quadratic after embedding. It is clear that beginning with
a quadratic singular Lagrangian (which is the case for most
interesting models) a number of linear primary (as well as
secondary) constraints and a quadratic Hamiltonian emerge.
Therefore, using our method, after embedding one finds a quadratic
embedded Hamiltonian together with some linear primary constraints
in the same number as the original model. This guarantees that one
can find the embedded singular (Wess-Zumino) Lagrangian, with the
primary constraints of the embedded model as the primary
constraints.

2)When a chain structure exists the number of independent gauge
parameters is much less than when we lack it. By "gauge
parameters" we mean arbitrary functions of time that appear in the
solutions of equations of motion. To be more precise, in the
traditional BFT method where we have ultimately the strongly
involutive constraints $\tau_\alpha$ and Hamiltonian $\tilde{H}$,
it is clear that the following function acts as the generating
function of gauge transformation
 \begin{equation}
 G=\sum\varepsilon^\alpha(t) \tau_\alpha(q,p,\eta).
 \label{g1}
 \end{equation}
As is apparent, here the number of gauge parameters
$\varepsilon^\alpha(t)$ is equal to the total number of
constraints.

However, in the presence of the chain structure it can be shown
that the number of gauge parameters is just equal to the number of
distinct chains \cite{ShirLoran}. There exist two equivalent
methods to construct gauge generating function
\cite{Henn,ShirShab}. In the special case, where all the
constraints are abelian and the terminating elements of each chain
commute strongly with Hamiltonian, the gauge generating function
can be written as
\begin{equation}
 G=\sum_{i=1}^m \sum_{\alpha=1}^{N_i}\
\left(-\frac{d}{dt}\right) ^{N_i-\alpha} \zeta_i(t)
\Phi_\alpha^{(i)}
 \label{b2}
 \end{equation}
where $\zeta_i(t)$ are infinitesimal arbitrary functions of time,
$m$ is the total number of first class chains and $N_i$ is the
length of the $i$-th chain.

It should be remembered that the generating function (\ref{g1})
gives the symmetries of the extended action, while $G$ in
(\ref{b2}) gives the symmetries of the total action. The former
formalism (i.e. the extended Hamiltonian), however, may be shown
to give the correct results for physical (non-gauged) observables.

Now it is instructive to apply the general idea discussed above to
a specific model. The bosonized chiral Schwinger model in $(1+1)$
dimensions with regularization parameter $a=1$ is described by the
Lagrangian density \cite{JacRaj,MitraRaj}:
 \begin{equation}
{\cal{L}}^N = \frac{1}{2} \partial_\mu \phi \partial^\mu
  \phi + (g^{\mu\nu} - \varepsilon^{\mu\nu}) \partial_\mu\phi
  A_\nu - \frac{1}{4} F_{\mu\nu} F^{\mu\nu} + \frac{1}{2} A_\mu
  A^\mu
 \label{c1}
 \end{equation}
in which $\phi$ is a scalar and $A_\mu$ is a vector field. There
is one second class chain including four second class constraints
as follows:
  \begin{equation}
  \begin{array}{l}
  \Phi_1 \equiv \pi_0 \approx 0 \\ \Phi_2\equiv
  E' + \phi' + \pi + A_1 \approx 0 \\ \Phi_3 \equiv E\approx 0 \\ \Phi_4 \equiv -\pi - \phi' - 2A_1 + A_0 \approx
  0
  \end{array}
  \label{c2}
  \end{equation}
where $\pi$, $\pi_0$ and $E$ are momenta conjugate to $\phi$,
$A_0$, and $A_1$ respectively. The canonical Hamiltonian density
corresponding to Eq. (\ref{c1}) is
  \begin{equation}
  {\cal{H}}^N_C = \frac{1}{2} \pi^2 + \frac{1}{2}\phi'^2 +
  \frac{1}{2}E^2 + EA'_0 + (\pi + \phi' + A_1)(A_1 - A_0).
  \label{c3}
  \end{equation}
Eqs. (\ref{c2}) represent a second class constrained system with
the algebra
 \begin{equation}
 \{\Phi_i(\textbf{x},t) , \Phi_j(\textbf{y},t)\}=\Delta_{ij}\delta(\textbf{x}-\textbf{y})
 \label{c4}
 \end{equation}
where
 \begin{equation}\Delta= \left(
 \begin{array}{cccc}
 0&0&0&-1 \\0&0&1&0 \\0&-1&0&2 \\ 1&0&-2&0
 \end{array}\right).
 \label{c5}
 \end{equation}
Let define four auxiliary fields $\eta^\alpha(x)$ with the algebra
given by $\omega=-\Delta$. As discussed above, the choice $\chi=1$
satisfy Eq. (\ref{a12}). Then following the instructions given in
Eqs. (\ref{a11}) and (\ref{a13}), the new set of first class
constraint are found to be
\begin{equation}
  \begin{array}{l}
  \tau_1 \equiv \pi_0 + \eta^1 \approx 0 \\ \tau_2\equiv
  E' + \phi' + \pi + A_1 +\eta^2 \approx 0 \\ \tau_3 \equiv E + \eta^3 \approx 0 \\
  \tau_4 \equiv -\pi - \phi' - 2A_1 + A_0 + \eta^4 \approx
  0.
  \end{array}
  \label{c7}
  \end{equation}
From Eqs. (\ref{a18}-\ref{a20}) the embedded Hamiltonian which
preserves the chain structure (except for the last element) is
 \begin{equation}
 \tilde{\cal{H}}={\cal{H}}_c^N+{\cal{H}}^{(1)}+{\cal{H}}^{(2)}
 \label{c8}
 \end{equation}
where
 \begin{equation}
 {\cal{H}}^{(1)}=-\eta^1(\phi''+\pi'+2A_1'+2E)
 \label{c9}
 \end{equation}
  \begin{equation}
 {\cal{H}}^{(2)}=\eta^{1'}\eta^2+\eta^2\eta^4+\eta^2\eta^2-\eta^1\eta^{1''}-2\eta^1\eta^3-\frac{1}{2}\eta^3\eta^3.
 \label{c10}
 \end{equation}
Here it can be checked that the constraints in Eqs. (\ref{c7})
satisfy the chain structure relation (\ref{b1}) where $\tau_1$ is
the primary constraint. The chiral Schwinger model has also been
discussed in \cite{MonemShir} where the set of first class
constraints are the same as (\ref{c7}) while the embedded
Hamiltonian $\tilde{H}$ is different. It is worth nothing that
this Hamiltonian does not differ from that of the strongly
involutive formulation by the mere addition of a suitable
combination of the first class constraints.

 According to Eq. (\ref{b2}) the gauge generating
function written in terms of just one infinitesimal gauge
parameter $\zeta(x,t)$ is
 \begin{equation}
 G=\int(-\zeta
 \tau_4+\dot{\zeta}\tau_3-\ddot{\zeta}\tau_2+\dot{\ddot{\zeta}}\tau_1)dx.
 \label{c11}
 \end{equation}
The infinitesimal gauge variations of the original and auxiliary
fields generated by $G$ are as follows
 \begin{equation}
 \begin{array}{l}
\delta \phi=\zeta-\ddot{\zeta} \hspace{2cm}\delta A_0= \dot{\ddot{\zeta}} \\
 \delta A_1= \partial \ddot{\zeta} + \dot{\zeta} \hspace{1.2cm}
 \delta \pi=\partial \zeta-\partial \ddot{\zeta}  \\ \delta \pi_0=\zeta  \hspace{2.2cm} \delta
 E=\ddot{\zeta}-2\zeta  \\ \delta \eta^1 =-\zeta \hspace{2.4cm} \delta \eta^2 =- \dot{\zeta} \\
 \delta\eta^3=2\zeta -\ddot{\zeta} \hspace{1.4cm} \delta
 \eta^4=2\dot{\zeta}-\dot{\ddot{\zeta}}.
 \end{array}
 \label{c111}
  \end{equation}
It can be directly checked that the total action is invariant
under these variations.

We can redefine the auxiliary fields $\eta^1$ to $\eta^4$ into two
fields $\eta, \xi$ and their canonical momenta $\pi_\eta$ and
$\pi_\xi$ as follows
 \begin{equation}
 \begin{array}{l}
\eta=\eta^1 \hspace{2cm} \pi_\eta=\eta^4+2\eta^2 \\ \xi=\eta^3
\hspace{2cm} \pi_\xi=\eta^2.
 \end{array}
 \label{g2}
  \end{equation}
  Fortunately all the terms in Hamiltonian (\ref{c8}) are
  quadratic. This make it easy to reconstruct the following
  Lagrangian

 \begin{equation}
 {\cal L}={\cal L}^N + \eta (\phi''+\pi^1+2A^1+2E)+
 (\dot{\eta}^2-\eta'^2-\dot{\eta}\eta'+\dot{\eta}\dot{\xi}+2\eta\xi+\frac{1}{2}\xi^2)-\eta\dot{A}^0.
 \label{g3}
  \end{equation}
  The first term is the original Lagrangian (\ref{c1}), the
  second and third terms are due to ${\cal H}^{(1)}$ and ${\cal H}^{(2)}$ in Eqs. (\ref{c9}) and (\ref{c10}) respectively and the last term $(-\eta\dot{A}^0)$ is the crucial term
  which converts the primary constraint $\pi_0$ into $\pi_0+\eta$
  of the embedded system. One can directly check that beginning
  with the Lagrangian (\ref{g3}) the first class system given by
  Hamiltonian $\tilde{H}$ in Eq. (\ref{c8}) and
  $\tau_1$ to $\tau_4$ in Eq. (\ref{c7}) would be obtained.

\section*{Acknowledgment}
The authors thank referee for giving essential comments.

\end{document}